\documentclass[preprint]{aa} % for the letters 
\usepackage{graphicx}
\usepackage{txfonts}
\begin{document}

   \title{Coherence loss in phase-referenced VLBI observations}

   \author{I. Mart\'{i}-Vidal
          \inst{1}\thanks{Alexander von Humboldt Fellow}
          \and
          E. Ros\inst{2,1}
          \and
          M.A. P\'erez-Torres\inst{3}
          \and
          J.~C. Guirado\inst{2}
          \and
          S. Jim\'enez-Monferrer\inst{2}
          \and
          J.~M. Marcaide\inst{2}
          }

   \institute{Max-Planck-Institut f\"ur Radioastronomie, Auf dem H\"ugel 69,
              D-53121 Bonn (Germany)\\
              \email{imartiv@mpifr.de}
              \and
              Departament d'Astronomia i Astrof\'{\i}sica, 
              Universitat de Val\`encia, E-46100 Burjassot, Val\`encia (Spain)
              \and
              Instituto de Astrof\'{\i}sica de Andaluc\'{\i}a, CSIC, 
              Apdo. Correos 2004, E-08071 Granada (Spain)
             }

   \date{Submitted on 5.02.2010. Accepted on 11.03.2010.}
 
  \abstract
   {Phase-referencing is a standard calibration technique in radio interferometry, particularly suited for the
detection of weak sources close to the sensitivity limits of the interferometers. However, effects from a 
changing atmosphere and inaccuracies in the correlator model may affect the phase-referenced images, and lead
to wrong estimates of source flux densities and positions. A systematic observational study of signal 
decoherence in phase-referencing and its effects in the image plane has not been performed yet.}
   {We systematically studied how the signal coherence in Very-Long-Baseline-Interferometry (VLBI) 
observations is affected by a phase-reference calibration at different frequencies and for different 
calibrator-to-target separations. The results obtained should be of interest for a correct interpretation 
of many phase-referenced observations with VLBI.}
   {We observed a set of 13 strong sources (the S5 polar cap sample) at 8.4 and 15 GHz in 
phase-reference mode with 32 different calibrator/target combinations spanning angular separations between
1.5 and 20.5 degrees. We obtained phase-referenced images and studied how the 
dynamic range and peak flux-density depend on observing frequency and source separation.}
   {We obtained dynamic ranges and peak flux densities of the 
phase-referenced images as a function of frequency and separation from the calibrator. We
compared our results with models and phenomenological equations previously reported.}
   {The dynamic range of the phase-referenced images is strongly limited by the atmosphere at all 
frequencies and for all source separations. The limiting dynamic range is inversely proportional 
to the sine of the calibrator-to-target separation. Not surpriseingly, we also find that the 
peak flux densities decrease with source separation, relative to those obtained from the 
self-calibrated images.}

   \keywords{Techniques: interferometric -- Atmospheric effects}

   \maketitle
%________________________________________________________________

\section{Introduction}

Phase-referencing is a standard calibration technique in radio interferometry. 
It allows the detection of a weak source (target source) by using quasi-simultaneous 
observations of a closeby strong source (calibrator) (see e.g., Ros \cite{ros05} 
and references therein). This technique also allows the user to recover the position of the 
target source relative to that of the calibrator; a position otherwise lost by the use of 
closure phases in the imaging. Basically, phase-referencing consists of 
estimating the antenna-based complex gains with the calibrator fringes,
time-interpolating these gains to the observations of the target, and
calibrating the visibilities of the target with the interpolated gains. 
Therefore, it is assumed that for each antenna the optical paths of 
the signals from both sources are similar. However, atmospheric turbulences 
and/or inaccuracies in the correlator geometrical model may 
introduce errors in the estimates of the optical paths of the signals 
and severely affect the phase-referencing. These errors can be partially corrected
by applying the self-calibration algorithm (see, e.g., Readhead \& Wilkinson 
\cite{Readhead1978}) after 
phase-referencing. However, self-calibration, specially on observations 
of weak sources, may affect the resulting images in undesirable ways (see, e.g., 
Mart\'i-Vidal \& Marcaide \cite{mar08b}).   

The correlator model includes contributions from the dry troposphere, the Earth 
orientation parameters, gain corrections for the sampling, and feed rotation of 
the alt-azimuthal mounts of the antennae.  An imperfect modelling in any of
these contributions and the loss of coherence of the radio waves within the 
time elapsed between consecutive observations of a given source, have an impact 
on the quality of phase-referencing. Some authors have stated that the loss 
of signal coherence in phase-referencing is linearly dependent on the separation 
between the calibrator source and the target source (Beasley \& Conway \cite{bea95}). 
However, in a recent publication Mart\'{\i}-Vidal et al. (\cite{mar10}) suggest a 
phenomenological model different from that of Beasley \& Conway (\cite{bea95}),
based on Monte Carlo simulations of atmospheric turbulences.

To empirically establish this dependence, we need to compare phase-referenced images to 
those obtained from the self-calibrated visibilities (i.e., the images obtained by applying 
the {\em optimum} phase gains; those that maximize the signal coherence of the target sources). 
For this purpose, both calibrator and target must be strong enough to generate fringes 
with a signal-to-noise ratio (SNR) high enough to allow for an accurate estimate of 
the phase, delay, and rate of the fringe peaks, thus avoiding the bias effects related 
to self-calibration of weak signals (e.g., Mart\'i-Vidal \& Marcaide \cite{mar08b}).

The S5 polar cap sample (a subset of 13 sources from the S5 survey, 
see K\"uhr et al. \cite{Kuhr1981}, Eckart et al. \cite{Eckart1986}) is an ideal
set of sources to perform such a study. The spectra of these sources are relatively 
flat at radio wavelenghts and their flux densities range from hundreds of mJy to 
several Jy. Therefore it is possible to study the loss of phase coherence as a 
function of observing frequency and source separation. We observed the S5 polar cap 
sample at 8.4\,GHz and 15\,GHz. These observations are part of a large 
astrometry programme (Ros et al. \cite{ros01}, P\'erez-Torres et al. \cite{per04},
Mart\'{\i}-Vidal et al. \cite{mar08}, and Jim\'enez-Monferrer et al.
in preparation) and were performed in phase-reference mode with many different 
calibrator/target combinations. From these observations, we studied the 
loss of signal coherence in phase-referenced observations by comparing 
phase-referenced and self-calibrated images for all possible 
calibrator/target combinations allowed by the observations. 

In the next section we describe our observations and the 
process of data calibration and reduction. In Sect. \ref{sec:results} we 
report on the results obtained. In Sect. \ref{sec:conclusions} we summarize 
our conclusions.

%__________________________________________________________________

\section{Observations and data reduction \label{sec:obs}}

We observed the 13 sources of the S5 polar cap sample with the complete 
Very Long Baseline Array (VLBA) at 8.4\,GHz on 2001 February 3 and at 15\,GHz 
on 2000 June 15. At each epoch, 8 bands of 8\,MHz bandwidth each were recorded, 
obtaining a synthesized bandwidth of 64\,MHz 
in one polarization (LCP). All three epochs consisted of 24 hr of 
observations. The observations took place in phase-referenced mode with 
different subsets of three or four radio sources in different duty cycles 
(see Mart\'{\i}-Vidal et al. \cite{mar08} for more details on the 
observing schedule). The sources of each subset were observed cyclically for 
about two hours. Each radio source was observed a total of about four hours. 
Data were cross-correlated at the Array Operation 
Center of the National Radio Astronomy Observatory (NRAO). Details on the 
source images at 8.4\,GHz and 15\,GHz can be found in Ros et al. 
(\cite{ros01}) and P\'erez-Torres et al. (\cite{per04}), respectively. 
The results of the high-precision astrometry analysis at 15\,GHz can be found 
in Mart\'{\i}-Vidal et al. (\cite{mar08}). Those at 
8.4\,GHz will be published elsewhere (Jim\'enez-Monferrer et al. in preparation).

The data calibration and reduction was performed using {\em ParselTongue} 
(Kettenis et al. \cite{Kettenis2006}), a Python interface to the NRAO 
Astronomical Image Processing System ({\sc aips}). We generated a script 
in ParselTongue to automatize the calibration and imaging of the 
phase-referenced images between all source pairs that were 
observed within the same duty cycles. A total of 32 source pairs were 
obtained, covering a range of source separations between 1.5 and 20.5 
degrees. 
We checked the quality of the automated calibration and
imaging by re-generating manually phased-referenced images of some 
pairs of sources and comparing them to the automated images. The results 
obtained were compatible within 0.1$\sigma$. Below we 
summarize the process of calibration followed in our analysis.

\begin{itemize}

\item {\bf Step 1.}
A correction of the the Earth Orientation Parameters 
(EOP) as estimated by the United States Naval Observatory (USNO) 
was applied to the data. 

\item {\bf Step 2.}
The visibility phases were aligned between the sub-bands, through 
the whole 64\,MHz band for all sources and times by fringe-fitting the 
sub-band delays of a selected scan with high-quality fringes
and applying the estimated delays and phases to all visibilities.

\item {\bf Step 3.}
A second fringe fitting, using now the delays determined from the 
whole band, 
provided new phase corrections for all the observations. The visibility
amplitude calibration was performed with the system temperatures
and gain curves from each antenna. 

\item {\bf Step 4.}
The calibrated data were exported into the programme {\sc difmap} (Shepherd 
et al. \cite{Shepherd1995}) to obtain images of all sources. The CLEAN 
algorithm and several iterations of phase and gain self-calibration were 
applied to each source until high-quality images (with residuals close to 
the thermal noise) were obtained.

\item {\bf Step 5.}
These images were imported back into {\sc aips} for a second fringe 
fitting, now taking into account the contribution of the source 
structures in the estimates of the residual delays and phases. The 
amplitude calibration was also refined by estimating the amplitude 
gains based on the source structures (one gain solution for all 
antennae every 10 minutes).

\end{itemize}

Once the data were calibrated as described above, we proceeded with 
the analysis.
For each pair of sources (A and B), four sets of data were generated.
On the one hand, the self-solutions of A (B) in the time range when 
this source was observed in the same duty cycles as B (A). On the other
hand, the set of visibilities of A (B) phase-referenced to B (A), 
obviously for the same time range as that of the corresponding self-calibrated
data sets. Each of these data sets was used to generate an image using
natural weighting of the visibilities (to optimize the sensitivity) and the 
same CLEAN windows that were used to obtain the images in Step 4 
of the data calibration procedure. For the phase-referenced images, the 
position of the peak flux-density was first estimated and the CLEAN windows 
were accordingly shifted before CLEANing.

\section{Results and discussion}
\label{sec:results}

\subsection{Dynamic range}

The dynamic ranges of the images obtained as described in the previous section
were computed as the peak flux-density per unit beam divided by the
root-mean-square (rms) of the image residuals (i.e., after subtracting the CLEAN
model). The peak flux densities of the images obtained from the self-calibrated 
visibilities range between 0.16 and 2.60\,Jy at 8.4\,GHz and between 0.16 and 
1.86\,Jy at 15\,GHz. The typical rms of these images are 1\,mJy\,beam$^{-1}$ at 
8.4\,GHz and 2\,mJy\,beam$^{-1}$ at 15\,GHz. 

\begin{figure}
\centering
\includegraphics[width=9cm]{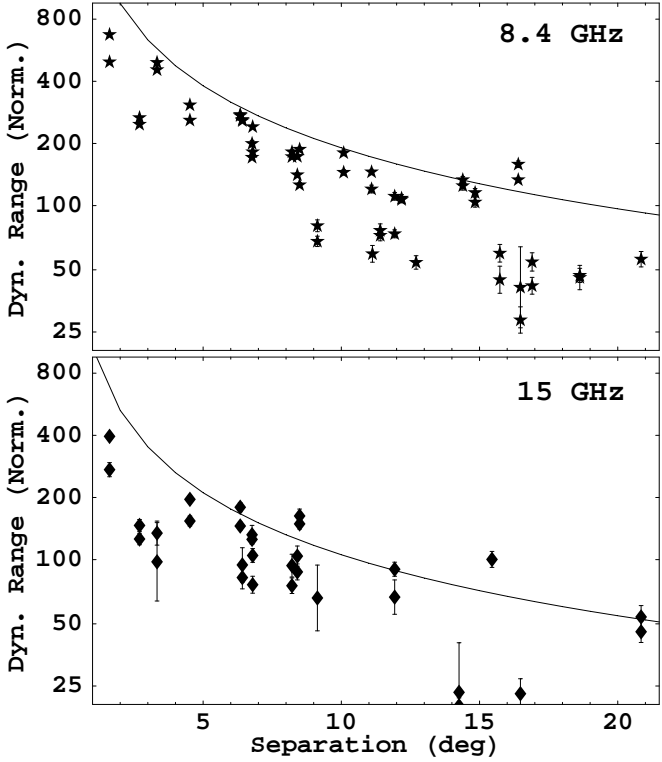}
\caption{Dynamic ranges (normalized to an observing time of 10 hours) of the 
phase-referenced images as a function of distance to the calibrators. The error 
bars are proportional to the flux densities of the calibrators. Lines represent 
a model of the maximum achievable dynamic range (Eq. \ref{DL1}).}
\label{DynRPh}
\end{figure}

\begin{figure}
\centering
\includegraphics[width=9cm]{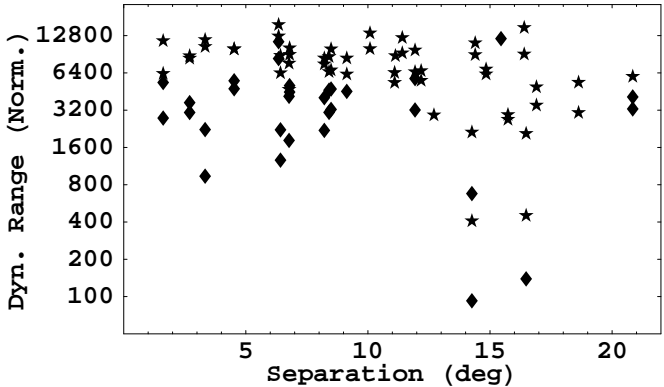}
\caption{Dynamic ranges (normalized to an observing time of 10 hours) of the 
images obtained from self-calibrated visibilities. Stars are data at 8.4\,GHz 
and diamonds at 15\,GHz. This figure can be compared to Fig. \ref{DynRPh} to 
see the effect of coherence losses in the phase-referenced images.}
\label{DynRSf}
\end{figure}

A total of 64 phase-referenced images were obtained at each frequency. These 
are two images for each pair of sources (i.e., image of source A phase-referenced 
to B and image of source B phase-referenced to A). We then 
discarded the images with dynamic ranges lower than 10, because in these cases 
there is a high chance of confusion of the source with a spurious noise peak. 
Applying this cutoff to the dynamic ranges, 53 phase-referenced images were 
left at 8.4\,GHz and 31 images at 15\,GHz. The dynamic ranges of the 
phase-referenced images are typically a factor $\sim40$ smaller than those 
obtained from the self-calibrated visibilities (see Figs. \ref{DynRPh} and \ref{DynRSf}). 
A first conclusion is that {\em the loss of phase coherence strongly affects the dynamic 
range of the phase-referenced images, regardless of the calibrator-to-target 
separation.}

A similar conclusion was reported by Mart\'i-Vidal et al. (\cite{mar10})
based on Monte Carlo simultations of phase-referenced observations under a 
turbulent atmosphere. These authors modelled the dynamic range of a 
phase-referenced image considering the addition in quadrature of two sources of 
noise; one due to the thermal noise of the receiving system, $\sigma_{th}$,
and the other due to the atmosphere. $\sigma_{at}$. The latter was assumed 
to be equal to a given percentage of the source flux-density, i.e. $\sigma_{at} = f_{at}\,S$.
In this expression, $S$ is the flux density and $f_{at}$ is a factor that depends 
on the calibrator-to-target separation, $\theta$, the observing frequency, 
$\nu$, and the on-source observing time $\Delta t$. We notice that a similar expression 
is usually employed in the estimate of the (instrumental) systematic amplitude-calibration 
errors in VLA and VLBI observations. However, in our case the factor $f_{at}$ depends 
on the calibrator-to-target separation and on the total on-source observing time, while 
the factors for the estimate of systematic amplitude-calibration errors in VLA and VLBI 
observations only depend on the observing frequency. The dynamic range is thus

\begin{equation}
D = \frac{S}{\sqrt{\sigma_{th}^2 + f_{at}^2\,S^2}} .
\label{DPR}
\end{equation}

The dynamic range of a phase-referenced image is limited to a given value, 
$D_l$,

\begin{equation}
D_l = \frac{1}{f_{at}} .
\label{DL1}
\end{equation}

This limit is independent of both $S$ and the sensitivity of the array. It is 
achieved when the flux density of the 
target source is much higher than the thermal noise of the 
interferometer. The limiting dynamic range only depends on $\theta$, 
$\nu$, and $\Delta t$, and it can be several orders of 
magnitude smaller than the corresponding dynamic range due to the 
thermal noise of the receiving system, indicating that the atmosphere
strongly limits the sensitivity of the observations. 
Mart\'i-Vidal et al. (\cite{mar10}) propose a phenomenological model 
for $f_{at}$ based on their Monte Carlo simulations (see their Sect. 
5.3). This model takes the form

\begin{equation}
f_{at} = K \frac{\nu}{\sqrt{\Delta t} \sin{(\theta)}} ,
\label{FAT}
\end{equation}

\noindent where $K$ is a constant to be determined. 

We can compare this 
phenomenological model to our observations by plotting the dynamic
range of all our phase-referenced images as a function of 
calibrator-to-target separation. The images obtained for 
each source pair correspond to data with slightly different 
on-source observing times. Therefore, we corrected (i.e., we normalized) 
the dynamic ranges of all the images by applying the factor 
$\sqrt{N_{10}/N_{v}}$, where $N_{v}$ is the number of visibilities 
for each image and $N_{10}$ is the number of visibilities corresponding 
to a on-source observing time of 10 hours with the VLBA. We show in 
Fig. \ref{DynRPh} the 
normalized dynamic ranges and the model resulting from
Eq. \ref{FAT}. We find that for 
$K \sim 12.4$\,hr$^{0.5}$\,GHz$^{-1}$ the 
model predicts the limiting dynamic ranges obtained with the VLBA at 
both frequencies, although the results at 8.4\,GHz are of higher quality;
the results at 15\,GHz are noisier. The dynamic ranges 
(also normalized to an observing time of 10 hours) obtained from the 
self-calibrated images are shown in Fig. \ref{DynRSf} for a comparison 
with those obtained from phase-referencing. The error bars in Fig. \ref{DynRPh} 
are set to be proportional to the flux density of the calibrator.
This way the reader has information in a single figure on the quality of 
the phase-referenced images and the quality of the calibrator visibilities.

%We notice that the dynamic ranges of the 
%self-calibrated images for similar source separations can differ by 
%a factor of $6-7$, while the variation is much smaller for the 
%phase-referenced images (a factor of only $2-3$). This convergence of 
%dynamic ranges for similar source separations in the phase-referenced 
%images, regardless of the very different flux densities of the sources, 
%is an effect of the limiting dynamic range, $D_l$, and indicates that our 
%phase-referenced observations were performed in the limit of strong 
%sources (as it is, indeed, the case).

\subsection{Peak flux-density}

\begin{figure}
\centering
\includegraphics[width=9cm]{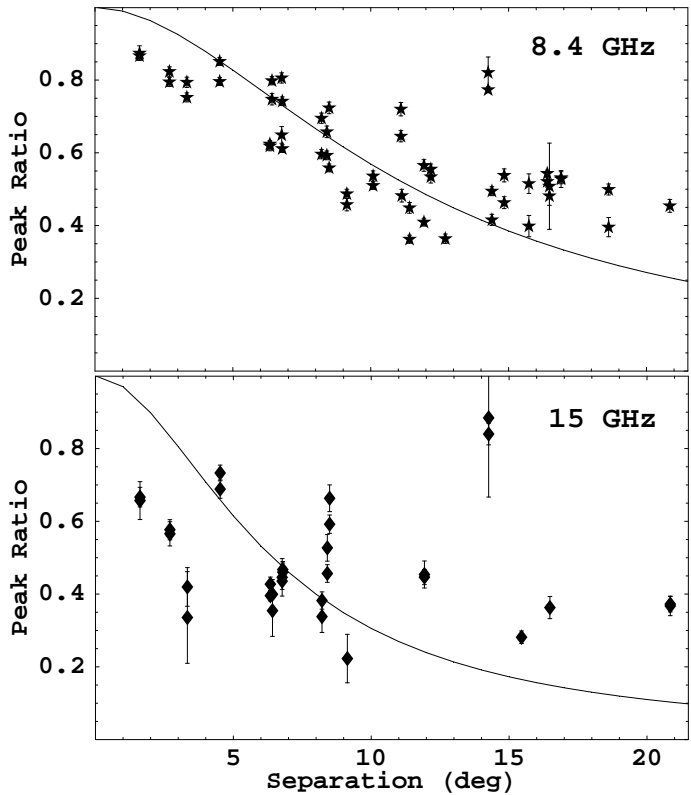}
\caption{Peak flux densities of the phase-referenced images normalized to the 
peak flux densities of the images obtained from the self-calibrated visibilities. 
Lines represent the model given by Eq. \ref{PEAKRATIO}.}
\label{PEAKRATIOFIG}
\end{figure}

We show in Fig. \ref{PEAKRATIOFIG} the peak flux densities of the 
phase-referenced images relative to those obtained from the images 
corresponding to the self-calibrated visibilities as a function of 
distance to the calibrator. The systematics in the loss of flux 
density is clear for the 8.4\,GHz data. The flux density recovered is
about 80\% of the real flux density for separations of $\sim5$ degrees,
and slowly decreases to 50\% for the largest separations. For the 
15\,GHz data, the scatter is larger and no robust conclusion can be 
obtained. However, a hint of saturation of the 
peak ratios around 40\% can be appreciated at large source separations 
if the points corresponding to a separation of 14.84 degrees (for which 
peak ratios as high as 80\% are obtained) are not considered (these 
points correspond to sources B1803+784 and B1150+812 phase-referenced 
to each other).

From the simulations reported in Mart\'i-Vidal et al. (\cite{mar10}),
the peak flux-density of a phase-referenced image, $P_{ph}$, relative 
to the real peak flux-density of the source, $P_{sf}$, can be estimated 
as

\begin{equation}
\frac{P_{ph}}{P_{sf}} = \frac{1}{1 + k_1\,f_{at}^{k_2}} ,
\label{PEAKRATIO}
\end{equation}

\noindent where $k_1$ is a constant to be determined. From the Monte 
Carlo simulations, the exponent $k_2$ takes the value $1.87 \pm 0.02$. 
In Fig. \ref{PEAKRATIOFIG} we also show the predictions of the model 
given by Eq. \ref{PEAKRATIO} for $k_1 \sim 63$. 
The model roughly predicts the behaviour of the 8.4\,GHz data, although 
for separations larger than $\sim15$ degrees, the observed peak ratios 
clearly saturate at $\sim0.5$, while the model predictions monotonically 
decrease. A similar conclusion is obtained for the 15\,GHz data, where 
a saturation around 0.4 is appreciated. The saturation in the ratio of 
flux-density peaks at large calibrator-to-target separations is 
not modelled using Eq. \ref{PEAKRATIO} and could be due to the saturation of
the power-spectrum of the tropospheric turbulences at large scales (see, e.g., 
Thomson, Moran, \& Swenson \cite{TMS91}), which was not considered in the 
simulations reported in Mart\'i-Vidal et al. (\cite{mar10}). This
saturation of the power spectrum of the turbulences would stabilize the 
phase difference between target and calibrator for large separations and 
therefore enhance the signal coherence (and peak flux-density) of the 
target source.

\section{Conclusions}
\label{sec:conclusions}

We report how phase referencing affects the signal coherence 
(and the fidelity of the images) in VLBI 
observations at different frequencies (8.4\,GHz and 15\,GHz) and for different 
calibrator-to-target separations (from 1.5 to 20.5 degrees).
We determined the loss of dynamic range and peak flux-density of the 
phase-referenced images and compared the results with the model predictions
given in Mart\'i-Vidal et al. (\cite{mar10}). 

The dynamic range of the
phase-referenced images is strongly limited by the atmosphere at all frequencies
and for all calibrator-to-target separations. The maximum achievable dynamic range
using the VLBA is given by Eqs. \ref{DL1} and \ref{FAT}, with $K \sim 12.4$ (if 
$\nu$ is given in GHz and $\Delta t$ in hours). If the target source is not strong
(as is usually the case), the thermal noise of the receiving system cannot be 
ignored and the dynamic range should be estimated with Eqs. 
\ref{DPR} and \ref{FAT}.

The flux-density (computed as the peak flux 
density of the phase-referenced image in units of the real peak flux-density of the
source) decreases as the separation to the calibrator increases and is given by 
Eqs. \ref{PEAKRATIO} and \ref{FAT}, with $k_1 \sim 63$ and 
$k_2 \sim 1.87$. This model roughly predicts the peak ratios observed at 8.4\,GHz 
for calibrator-to-target separations below $\sim 15$ degrees (the results at 
15\,GHz are too noisy to reach a robust conclusion). For separations larger 
than 15 degrees, the observed peak ratios are higher than the model predictions,
possibly due to a saturation in the power spectrum of the tropospheric turbulences
at large scales, which was not considered in Mart\'i-Vidal et al. 
(\cite{mar10}). 

It is remarkable that for relatively small separations 
(below 5 degrees), which are typical in many phase-referencing observations, the 
flux-density loss can be as large as 20\% at 8.4\,GHz and $30-40$\% at 15\,GHz 
(and even larger at higher frequencies, according to Eq. \ref{PEAKRATIO}). It must
be also taken into account that the phase-referenced observations here reported 
were performed under good weather conditions and when the sources were close to their 
transits at nearly all stations (except Mauna Kea and St. Croix, see Mart\'i-Vidal et 
al. \cite{mar08} for details on the observing schedule). Therefore even larger 
flux-density losses and lower dynamic ranges may be obtained when the 
observing conditions are far from optimal.

However, we must also notice that the typical calibrator-to-calibrator 
cycle times in our observations were about $120-180$ seconds. For observations at 
8.4\,GHz, these cycle times are of the order of (and slightly shorter than) the 
coherence time of the signal (which takes typical values of $180-240$ seconds in VLBI 
observations). Therefore it is not expected that a changing atmosphere may have 
introduced strong effects in the coherence of the phase-referenced visibilities at 
8.4\,GHz. At 15\,GHz, the shorter coherence time (which takes typical 
values of $100-140$ seconds in VLBI observations) might be short enough to imply an 
additional phase degradation in the phase-referenced visibilities due to a changing 
atmosphere. Nevertheless, this issue is unlikely to significantly affect the results 
of our analysis, because the time evolution of the self-solutions of the 
antenna-based phase gains at 15\,GHz are smooth, which indicates that the weather 
conditions were good enough to allow for a well-behaved phase connection between 
contiguous scans of each source.

\begin{acknowledgements}

IMV is a fellow of the Alexander von Humboldt Foundation in Germany.
The Very Long Baseline Array is operated by the
USA National Radio Astronomy Observatory, which is a facility of 
the National Science Foundation operated under cooperative agreement 
by Associated Universities, Inc. 
Partial support was obtained by Generalitat Valenciana (Prometeo 2009 P104)
and from Spanish grant AYA 2005-08561-C03-02. MAPT acknowledges support by 
the MEC through grant AYA 2006-14986-C02-01, and by the Consejer\'ia de
Innovaci\'on, Ciencia y Empresa of Junta de Andaluc\'ia through grants 
FQM-1747 and TIC-126.

\end{acknowledgements}

\end{document}